\documentclass[showpacs,amsmath,amssymb]{revtex4}

\usepackage{latexsym}
\usepackage{amssymb}
\usepackage{graphics}
\usepackage{graphicx}% Include figure file
\usepackage{dcolumn}% Align table columns on decimal point
\usepackage{bm}% bold math

\begin{document}

\preprint{APS/123-QED}

\title{Effect of an electric field on a floating lipid bilayer: a neutron reflectivity study}
\author{Sigol\`ene Lecuyer}
\affiliation{Institut Charles Sadron, UPR 22, CNRS-Universit\'e Louis Pasteur, 6 rue Boussingault, F-67083 Strasbourg
Cedex, France}
\author{Giovanna Fragneto}
\affiliation{Institut Laue-Langevin, 6 rue Jules Horowitz, B.P. 156, 38042
Grenoble Cedex, France}
\author{Thierry Charitat}
\affiliation{Institut Charles Sadron, UPR 22, CNRS-Universit\'e Louis Pasteur, 6 rue Boussingault, F-67083 Strasbourg
Cedex, France}
% etc
%\thanks{\emph{Present address:}{CNRS - UPR 22, Universit\'e Louis Pasteur}
%date{Received: date / Revised version: date}
\begin{abstract}
We present here a neutron reflectivity study of the influence of an alternative electric field on a supported phospholipid double bilayer. We report for the first time a reproducible increase of the fluctuation amplitude leading to the complete unbinding of the floating bilayer. Results are in good agreement with a semi-quantitative interpretation in terms of negative electrostatic surface tension.
\end{abstract}


\maketitle

\section{Introduction}
\label{intro}

In an aqueous environment, phospholipids self-assemble into bilayers; these 2D systems have been recently extensively studied, as they display complex peculiar behaviours \cite{katsaras,liporevue}, and constitute the major component of the cell membrane \cite{mouritsen}. The strong influence of electric fields on lipid bilayers is interesting both from a fundamental point of view and for practical aspects, and has consequently received an important research effort. Large electric fields are also used to introduce macromolecules in vesicles or cells by 
formation of long-lived pores in the bilayer \cite{weaver1996,isambert98}, and also by endocytocis \cite{rosemberg(bioelec1997)}. Smaller alternative electric  fields are commonly used to destabilise lipid bilayers in order to form large unilamellar vesicles \cite{angelova86,ange2} in the so called {\it electroformation} technique. Recently Burgess et al. \cite{burgess(2004)} have studied the effect of a static electric field (surface charge) on the structure of supported bilayers.

\begin{figure}
\includegraphics[scale=0.7]{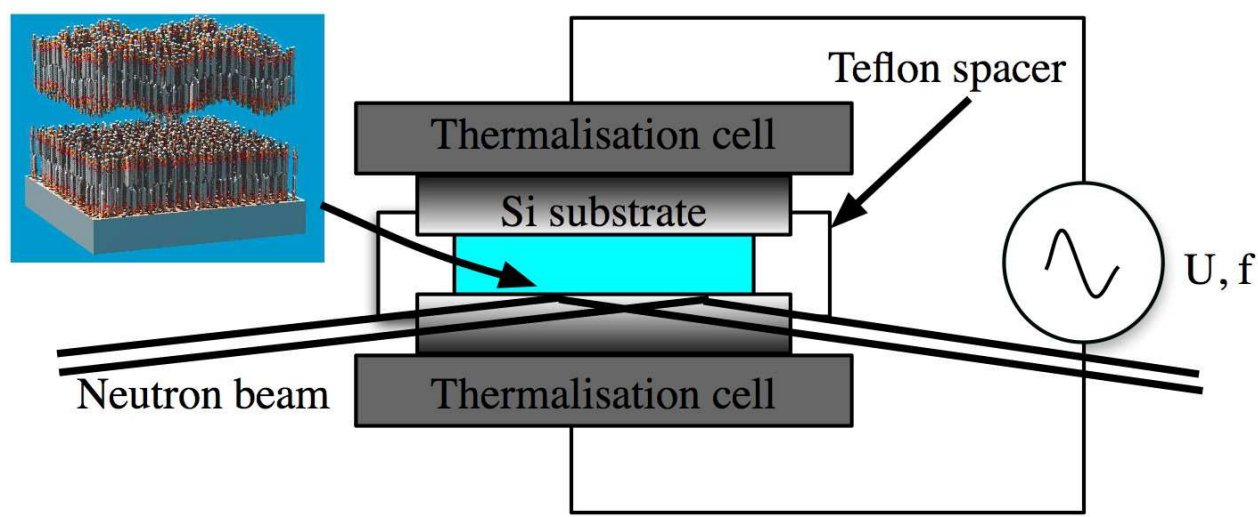}
\label{fig:1}
\caption{Experimental set-up: the two identical silicon substrates (10 mm-thick) are in thermal and electrical contact with thermalisation blocks. The PTFE spacer ensures a 5 mm separation between the two blocks.}
\end{figure}

Although it is a widely used technique, the mechanism involved in vesicle electroformation is not well understood yet. In a recent theoretical paper \cite{sens2002}, Sens and Isambert have suggested that a ''negative surface tension'' due to the electric  field induces an undulation instability of the bilayer, and that hydrodynamical effects select the fastest-destabilised mode. These predictions are difficult to test in classical electroformation experiments. As a matter of fact, the initial system is usually a rather disordered lamellar phase, obtained by hydration of a dry lipid film and, consequently, the result is very polydisperse in both size and degree of lamellar order. Recently, Constantin et al. \cite{salditt2005} have reported the first study of membrane destabilisation by an electric field on a more controlled system: a solid-supported fully-hydrated oriented lamellar phase (from 10 to 3000 bilayers). They observed for the first time individual peeling of the bilayers, but were not able to confirm or refute the existence of electrodynamical instability.

Because they are quite free of defects, double-bilayer systems are promising candidates to reach better understanding of the mechanism of vesicle formation in a well-defined geometry. Recent work has shown that they enable local study of the destabilisation of a single membrane \cite{lecuyer2006}. In this paper we focus on the first steps of the bilayer destabilisation by an electric field as detected by neutron reflectivity.

\section{Materials and Methods}
\label{sec:1}
\subsection{Sample preparation}
\label{sec:2}

Supported single and double-bilayers of DPPC and DSPC (1, 2-dipalmitoyl-sn-glycero-3-phosphocholine and 1, 2-distearoyl-sn-glycero-3-phosphocholine, Avanti Polar Lipids) molecules were prepared using the Langmuir-Blodgett and Langmuir-Schaefer techniques, as described in detail previously \cite{charitat1999}. 

Bilayers were deposited on silicon substrates (highly doped 2-10 $\Omega$.cm$^{-1}$, 5$\times$5$\times$1 cm$^{3}$), polished to produce a high quality crystal surface with r.m.s. roughness $< 0.2$ nm (SILTRONIX, Archamps, F).
After careful cleaning, the silicon blocks were made highly hydrophilic by a UV/ozone treatment as in \cite{charitat1999}. Lipid depositions were performed at a constant surface pressure of 40 mN/m, the Langmuir trough (NIMA, Warwick, UK) being thermalised at 20$\circ$C. Transfer ratios for the first layers were always high and of the same order as for samples described in \cite{charitat1999}.  A very precise adjustment of sample horizontality during the Schaefer dip, crucial for a good quality sample, led to transfer ratios always bigger than 0.96. After the deposition of the fourth layer, samples were constantly kept into water. They were closed while submerged in the trough, using a 5 mm-thick-PTFE spacer and a similar silicon substrate as a lid, and then placed into a water-regulated temperature chamber (see figure \ref{fig:1}). \\
Annealing was generally performed by progressively heating the sample to the fluid phase and then cooling it back to the gel phase. Then after careful caracterisation of the structure in the gel and the fluid phases, the adopted protocol consisted in applying to the supported fluid bilayer(s) an alternative electric field of fixed amplitude, decreasing gradually the frequency (from 200 Hz to 5 Hz). For each value of the frequency reflectivity data were collected. All measurements were made at 62$^\circ$C to ensure sample fluidity.

\subsection{Neutron reflectivity: basic principles and measurements}

Specular reflectivity, R(q), defined as the ratio between the specularly  reflected and incoming intensities of a neutron beam, is measured  as a function of the wave vector transfer, $q = 4\pi \sin{\theta}/\lambda$ perpendicular to the reflecting surface, where $\theta$ is the angle and $\lambda$ the wavelength of the incoming beam. R(q) is related to the scattering length  density across the interface, $\rho(z$), by the relation \cite{penfold}:
 					
\begin{equation}
R(q) = \frac{16 \pi^2}{q^2} \left| \tilde{\rho}(q) \right|^2
\label{r(q)}		
\end{equation}

where $|\tilde{\rho}(q)|$ is the one-dimensional Fourier Transform of $\rho(z)$.

Specular reflectivity allows the determination of the structure of matter perpendicular to a surface or an interface. Experiments are performed in reflection at grazing incidence. Samples must be planar, very flat and with roughness as small as possible. 
Data were collected at the High Flux Reactor (HFR) of the Institut-Laue Langevin (ILL, Grenoble, F) both on the small angle diffractometer D16 (preliminary measurements) and 
on the high flux reflectometer D17, designed to take advantage of both Time-Of-Flight (TOF) and monochromatic methods of measuring reflectivity (dual mode instrument) \cite{cubitt1,cubitt2}. In this study measurements were all taken with the instrument in the TOF mode by using a spread of wavelengths between 2 and 20 \AA \ at two incoming angles (typically 0.7$^{\circ}$ and 4$^{\circ}$) and with a resolution, defined by two choppers, of $\Delta t /t =1\%$ and 5$\%$, respectively, where $\Delta t$  is the neutron pulse duration. The beam was defined in the horizontal direction by a set of two slits, one just before the sample and one before a vertically focusing guide. 
Neutron reflectivity from the same sample was measured at different temperatures monitored with a thermocouple (equilibration time 25 
minutes, stability $<0.1^{\circ}$C, absolute precision 
$<0.3^{\circ}$C), in the water-regulated sample 
chamber already described in \cite{charitat1999}. 
 The useful q-range, before hitting the sample background, spanned from 0.007 to 0.24 \AA$^{-1}$ and was measured in about two hours, also after equilibration. The equilibration time varied depending on the step of temperature increase or decrease. For a step of 1$^{\circ}$C it was of about 15 minutes.

\subsection{Analysis of reflectivity data}

Specular reflectivity is a technique able to resolve the stacking of several slabs each one with a certain thickness and a certain scattering length density. Data analysis requires a good knowledge of the sample and here it was done via model fitting. Born and Wolf gave a general solution, the so called optical matrix method \cite{born}, to calculate the reflectivity from any number of parallel, homogeneous layers, which is particularly useful since any layered structure can be approximately described by dividing it into an arbitrary number of layers parallel to the interface, each having a uniform scattering length density. A reflectivity curve essentially reflects the sample density perpendicular to the substrate surface, or rather the square modulus of its Fourier transform. Since the phase is lost, data need to be fitted (e.g. with a slab model) to extract the density profile. In previous studies \cite{charitat1999}, multiple contrast neutron measurements have determined within \AA \ precision the profile of adsorbed and floating bilayers, in the gel and the fluid phases. Each bilayer is resolved into outer-inner-outer slabs: outer slabs = heads (phosphocholine and glycerol groups), inner slab = tails (hydrocarbon chains). A water film of thickness D$_w$  separates the adsorbed and floating bilayers. Taking into account the water layer between the adsorbed bilayer and the substrate and the silicon oxide layer, this leads to 9 slabs in total, each one having its own average thickness and scattering length density. Moreover, each interface between two slabs has a certain width, also called r.m.s. roughness. Since this enters into the density profile, it can be determined by fits of reflectivity curves, but it is not possible to determine whether it is a static intrinsic width or an average over the neutron beam's correlation length of temporal or spatial fluctuations.

\section{Results}

\subsection{Complete unbinding in a highly conductive solution}

\begin{figure}
\includegraphics[scale=0.6]{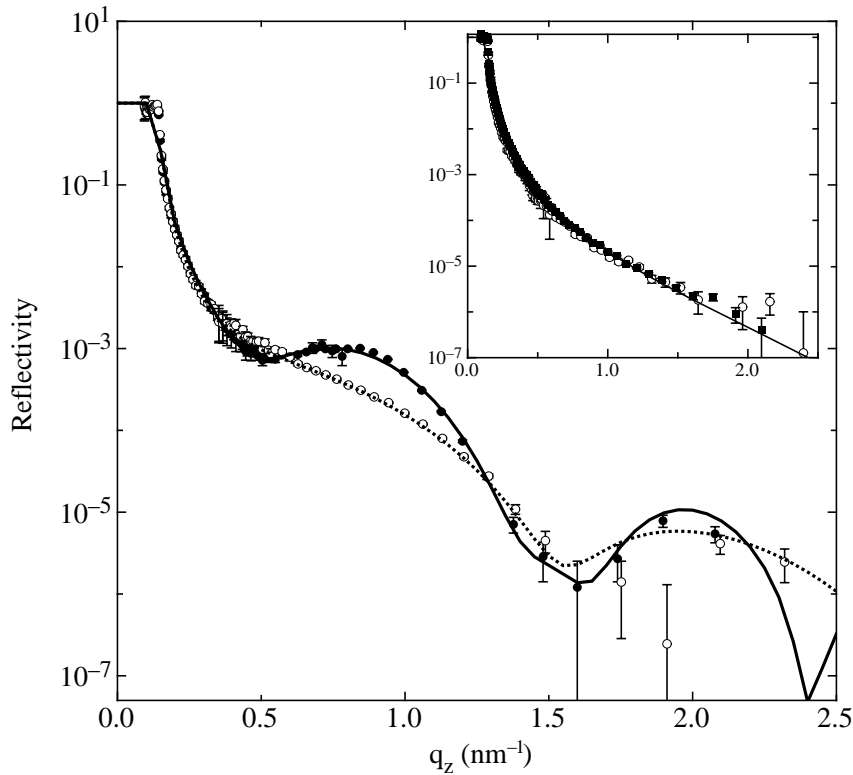}
\caption{Reflectivity curves: closed circles ($\bullet$) = double DSPC bilayer in CaCl$_2$ before applying electric field and best fit with a double bilayer model (full line); open circles ($\circ$) = after applying a (10 V, 10 Hz) field and best fit with a single bilayer model (dashed line). In insert, reflectivity curves for the bare substrate before and after the experiment (and best fit with a silicon oxyde layer model).}
\label{fig:2}
\end{figure}

The first result of this paper is the observation for the first time of the floating bilayer complete unbinding by application of an electric field. This effect was observed on two different samples under the same conditions: double bilayers of DSPC in a CaCl$_2$ in D$_2$O solution (0.35 mM.l$^{-1}$, measured conductivity of 16.62 $\mu$S.cm$^{-1}$) at low frequency (10 Hz) and for a voltage amplitude higher than 5 V. Figure \ref{fig:2} shows reflectivity data for one of these double bilayer DSPC samples before and after applying the electric field. The curves correspond to the best fits obtained with a 9-slab model and are in good agreement with previous multi-contrast neutron reflectivity experiments \cite{charitat1999}.

After the unbinding we still observed the presence of the first bilayer, without any major change in structure, but slightly shifted away from the substrate (figure \ref{fig:2}).  After removing the bilayer by cleaning the substrate we could confirm that the silicon oxide layer was unmodified and therefore that there was no formation of porous silicon oxide (figure \ref{fig:2} in insert). 
The adsorbed bilayer might possibly be removed by application of a lower frequency or/and higher voltage field, but this could not be tested as it could have induced significant oxidation of the silicon.
As a matter of fact, we had previously been able to determine that the formation of porous silicon oxide is very sudden and occurs when frequency becomes lower than 5 Hz in identical conditions (unpublished data). 

\begin{figure}
\includegraphics[scale=0.6]{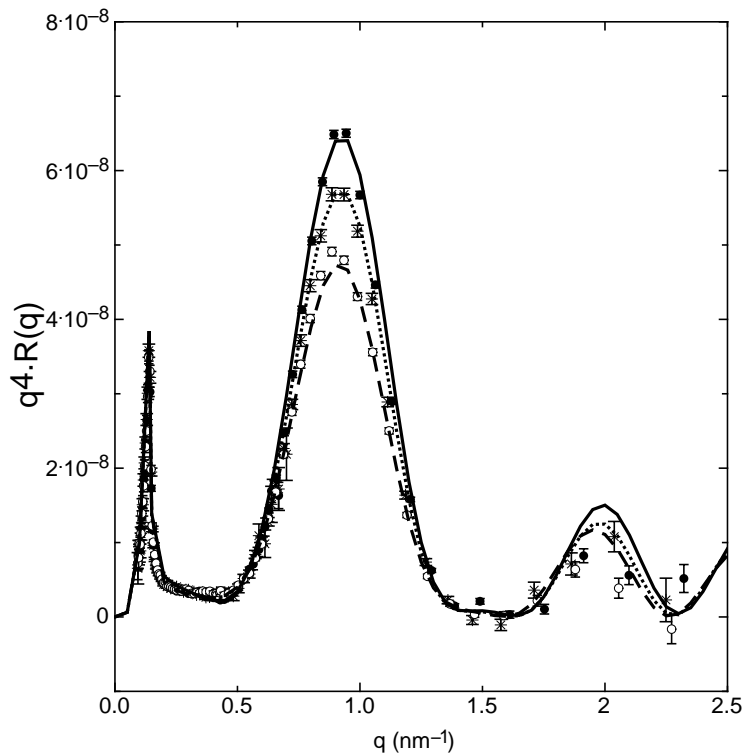}
\caption{Modification of the reflectivity curve with respect to the frequency of the electric field. A 5V alternative field is applied to a DSPC double bilayer in $D_2O$; full circles ($\bullet$) with f=100 Hz; stars ($\star$) with f=50 Hz; open circles ($\circ$) with f=10 Hz. The lines are the best fits obtained with the values of the roughness given in the text.}
\label{fig:3}
\end{figure}

\subsection{Progressive increase of fluctuations in a poorly conductive solution}

\subsubsection{Double bilayer}

As this complete unbinding of the floating bilayer is very sudden, we tried to observe previous steps of the destabilisation in a low conductivity electrolyte such as D$_2$O ($\chi_s^{-1} = 0.34~\mu$S.cm$^{-1}$). We performed experiments on a DSPC double bilayer in these conditions, and were able to observe a progressive modification of the reflectivity curves (figure \ref{fig:3}).\\

\begin{enumerate}

\item[$\bullet$]{At high frequencies (f $>$ 100 Hz), this modification appeared to be irreversible and involved mainly the first bilayer: the thin water layer ($\sim$0.3 nm, see table \ref{table1}) initially separating the substrate and the first bilayer disappears (or becomes smaller than our experimental resolution), while the amount of solvent in the bilayer increases from 2\% to 6\%. For this first confined bilayer, assuming that the first water layer behaves as a dielectric, both effects could originate from  the electrostatic pressure acting on the effective silicon oxyde-water layer-bilayer capacitance. Note that at the same time the second bilayer structure is unmodified.}

\item[$\bullet$]{At lower frequencies, modifications also involved the floating bilayer roughness and were partially reversible. It is very difficult to determine which structural parameter(s) is/are responsible for the subtle modifications observed without additional information. Nevertheless, as previously mentioned no modification of the silicon oxide layer was detected. The structure of the double bilayer can be well determined before applying the electric field \cite{charitat1999}, and the global structure of the bilayers (i.e. the slabs number and nature) seems to be unaffected by the electric field. 
It thus seemed physically reasonable to fit the reflectivity curves under electric field by varying mainly bilayers roughness and water layers thickness; average percentages of solvent within the bilayers were also allowed to vary, accounting for the induced formation of defects or pores in the membranes.}

\end{enumerate}

\begin{figure}[h]
\includegraphics[scale=0.6]{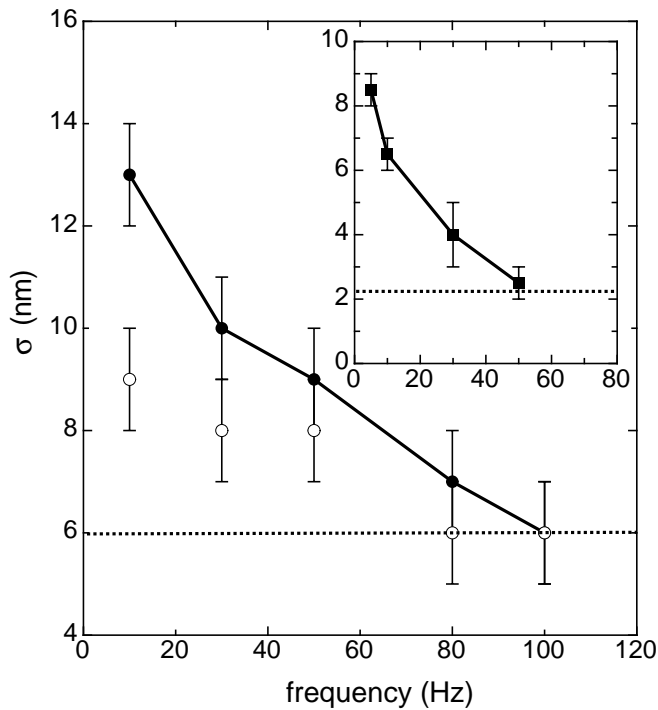}
\caption{Top bilayer roughness vs electric field frequency f at  constant field amplitude $V_{pp}/2 = 5 V$: ($\bullet$) DSPC double-bilayer in D$_2$O (in insert, ($\blacksquare$) same curve for a DSPC bilayer); ($\circ$) reversibility  test after each frequency (no electric field) for the double bilayer.}
\label{fig:4}
\end{figure}

Within this framework, the dominant effect we observed was a progressive increase of the floating (top) bilayer roughness, going from 0.6$\pm$1 to 1.3 $\pm$1 nm. Again it was not possible to go to lower frequencies or higher voltages to avoid substrate deterioration, and in the frequency-range studied no complete unbinding occured in D$_2$O.\\

This roughness increase seemed to be partially reversible, as shown on figure \ref{fig:4}. Reversibility was systematically checked in the low conductivity solution (D$_2$O); for beam time limitation reasons we were not able to check it in the CaCl$_2$ conductive solution.  A small increase of the thickness of water between the bilayers D$_{w,2}$, in the limit of the resolution of the experiment, was also observed. 
Finally, a progressive increase of the percentage of solvent in the first, substrate-bound bilayer was also observed, while the percentage of solvent in the floating bilayer remained constant and very low. Whether this corresponds to the formation of pores - made easier by the higher tension of the supported membrane, or of permanent holes due to electrochemical phenomena close to the substrate is unsure. However the latter seems more likely as this increase of the solvent content was apparently not reversible.\\ 
The irreversible effect on the first bilayer at frequencies $>$ 100 Hz, as well as the reversible effect at low frequencies, well described by an increase of the top bilayer roughness, were observed on different DSPC samples. Nevertheless, as we were not able to check carefully the stability of the oxide layer in these cases we do not report corresponding data in the present paper.\\

\subsubsection{Single bilayer}

In order to check the effect of the electric field on the first, substrate-bound bilayer, we performed experiments on a single supported DSPC bilayer. Results are shown in figure \ref{fig:4} (insert). They showed a progressive increase of the bilayer roughness, along with an increase of membrane-substrate distance. However no modification of the bilayer structure was seen, and the percentage of solvent remained very low ($\sim$2\%), in agreement with a good-quality sample. We can thus think that, in the case of the double-bilayer, the solvent increase in the first bilayer is correlated to the fact that it is confined close to the substrate by the second, fluctuating bilayer. This seems consistent with the fact that when the second bilayer unbinds, the first one relaxes away from the substrate.\\

\section{Discussion}

The results reported above clearly show that an electric field can induce a complete destabilisation of the top bilayer of a double-bilayer at low frequency ($\sim$ 5-10 Hz). 
This very abrupt destabilisation follows a reversible increase of the bilayer roughness. It is tempting to describe this roughness in terms of thermal fluctuations, and we intend to do so in this section.

\subsection{Bilayer fluctuations}

We consider the small thermal fluctuations of a single bilayer close to a substrate. Using common notations (see \cite{mecke2003} for example), we 
note $\vec{r}=(x,y)$ the in plane coordinates and $z$ the coordinate normal to the substrate and we call $u(\vec{r})=u(x,y)$ the membrane position. 
Classical Helfrich's Hamiltonian \cite{helfrich73} can then be written at first order expansion in membrane's curvature as~:

\begin{equation}
{\cal H}[u \left(\vec{r} \right)] = \int_{S} d^2 \vec{r}
\left[ U\left( u \left(\vec{r} \right) \right) +
\frac 12 \kappa \left(\Delta u \left(\vec{r}\right)\right)^2  + \gamma \left( \vec{\nabla} u\left( \vec{r} \right) \right)^2 \right] .
\label{hamiltonien}
\end{equation}

where $\kappa$ is the bending modulus of the bilayer, $\gamma$ its surface tension and $U$ the membrane-substrate interaction potential.
Using a quadratic approximation for $U\left( u \left(\vec{r} \right) \right)$ it is possible to apply the equipartition theorem for small fluctuations: 
\begin{eqnarray}
\left< |u_q|^2 \right> = \frac 1S \times \frac{k_B T}{U^{\prime\prime} + \gamma q^2 + \kappa q^4}
\label{uq2}
\end{eqnarray}

Note that this expression is only valid when the fluctuation amplitude is smaller than the mean distance to the substrate, which seems to be the case in our experiments. For large fluctuations one would have to take into account the reduction of entropy caused by the substrate, leading to the so-called Helfrich repulsion \cite{liporevue,mecke2003}.

In their recent paper, Sens and Isambert \cite{sens2002} describe the case of a free membrane far from a substrate ($U^{\prime \prime}=0$) and take into account the effect of hydrodynamics on the bilayer destabilisation. In particular, they showed that energy dissipation by solvent flow (viscosity $\eta$) and by monolayer-monolayer friction (friction coefficient $b_{fr}$) lead to the selection of a fastest-destabilised mode of wavelength $q^{\star}$:
\begin{equation}
q^{\star}=\frac{(\Gamma_{el}-\gamma) \eta}{\kappa b_{fr} \delta^2} \simeq 1~{\mu}m^{-1}
\end{equation} 
This length scale is in good agreement with the caracteristic size of vesicles prepared by electroformation. 

In our experimental study, the two bilayers are not free but interact with the substrate and the other bilayer. Bilayer equilibrium position without electric field is thus given by a competition between hydratation repulsive forces and van der Waals attractive ones, as described in ref.\cite{liporevue,mecke2003}. 

Using Eq. \ref{uq2}, it is possible to calculate the rms amplitude of fluctuations $\sigma^2$:
\begin{eqnarray}
\nonumber
\sigma^2 &=& \left< z \left(r\right)^2 \right> = \frac{1}{\left( 2 \pi \right)^2} \int d^2\vec{q} \left< |u_q|^2 \right>\\
&=& \frac{k_B T}{2\pi \sqrt{\Delta}} \log{\frac{q_2}{q_1}}
\label{sigmarms}
\end{eqnarray}

with $\Delta = \gamma^2 - 4 U^{\prime \prime} \kappa$, $q_1^2=\frac{\gamma - \sqrt{\Delta}}{2\kappa}$ and $q_2^2=\frac{\gamma + \sqrt{\Delta}}{2\kappa}$.

\begin{figure}[h]
\includegraphics[scale=0.6]{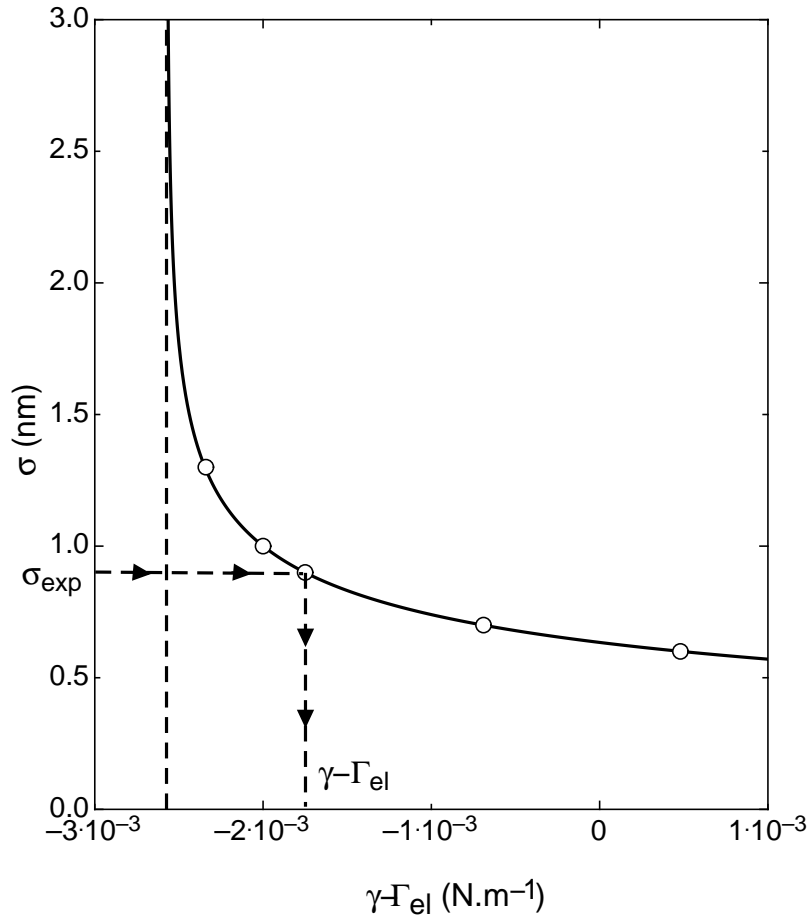}
\caption{Membrane rms fluctuation amplitude $\sigma$ as a function of global surface tension $\gamma - \Gamma_{el}$, deduced from equation  \protect\ref{sigmarms}. ($\circ$): values probed during the experiments.}
\label{fig:5}
\end{figure}

At the linear order, the effect of an electric field on the membrane can be expressed as a negative surface tension $-\Gamma_{el}$ (with $\Gamma_{el} > 0$). Following the notations of ref. \cite{sens2002}, $\Gamma_{el}$ can be written as:
\begin{equation}
\Gamma_{el} = \epsilon_m E_m ^2 \delta
\label{gammael}
\end{equation}
where $\epsilon_m$ is the membrane dielectric constant, $E_m$ is the local electric field and $\delta$ the bilayer thickness. This negative surface tension induces enhanced thermal fluctuations that can be described by replacing $\gamma$ by $\tilde{\gamma}=\gamma-\Gamma_{el}$ in equation \ref{uq2}. Figure \ref{fig:5} shows the variation of $\sigma$ as a function of the global surface tension $\gamma-\Gamma_{el}$; here we have taken $\kappa \simeq 40 k_B T$ and $U^{\prime \prime} \simeq 10^{13}$ J.m$^{-4}$,  which are classical values for a DSPC bilayer in the fluid phase \cite{daillant2005}. 

Membrane fluctuations undergo a sharp divergence for a finite negative value of $\tilde{\gamma_c} = - \sqrt{U^{\prime \prime}\kappa}$ which is the signature of bilayer destabilisation. This unbinding occurs in a very narrow surface tension range, which could explain the abrupt destabilisation
 observed in our experiments.

\subsection{Discussion in terms of electrostatic surface tension}

Let us now try to evaluate the electric surface tension equivalent to the field we have applied. It is possible to derive values of $\Gamma_{el}$ using equation  \ref{sigmarms}, if we know $\sigma$, $\gamma$, $\kappa$ and  $U^{\prime \prime}$. We measured $\sigma$ experimentally. Previous X-ray off-specular experiments \cite{daillant2005} have shown on very similar samples (a fluid DSPC bilayer deposited on an hybrid OTS-DSPC bilayer) that $\kappa \simeq$ 40 $k_B T$ and $\gamma \simeq$ 0.5 mN.m$^{-1}$. Using these experimental results we can fix $U^{\prime \prime}$ to obtain the correct experimental value of $\sigma$ measured without any electric field. Following this method we obtain $U^{\prime \prime} \simeq 10^{13}$ J.m$^{-4}$. This value is in good agreement with theoretical estimations using a van der Waals potential \cite{liporevue}. It is then possible by using equation \ref{sigmarms} to derive numerical values of $\Gamma_{el}$ for the different fields applied, which are reported on figure \ref{fig:5}.

Following ref. \cite{sens2002}, the electrostatic surface tension $\Gamma_{el}$ (Eq. \ref{gammael}) can be related to the electric field E applied to the whole cell by:
\begin{equation}
\Gamma_{el} \simeq \epsilon_m \left( \frac{\chi_m}{\chi_s} \right)^2 E^2 \delta
\label{estimgammael}
\end{equation}
where $\chi_m^{-1}$ and $\chi_s^{-1}$ are respectively the membrane and solvent conductivity. With $\epsilon_m \simeq 2\epsilon_0$, $\chi_s^{-1} = 0.34 \times 10^{-4}$ S.m$^{-1}$ for D$_2$O (measured experimental value) and $\Gamma_{el} \sim 10^{-3}$ N.m$^{-1}$ it leads to $\chi_m^{-1} \sim 3 \times 10^{-10}$ S.m$^{-1}$ (equivalent to a membrane resistance R$_m$ $\sim$ 0.2 M$\Omega$.cm$^2$), in good agreement with the value obtained by Purrucker et al. \cite{purrucker(2001)}. In the high conductivity solution (CaCl$_2$, $\chi_s^{-1} = 16 \times 10^{-4}$ S.m$^{-1}$), even if we found no value of the membrane conductivity reported in literature, it seems clear that we are well above the unbinding surface tension $\gamma_c$.

\begin{figure}[h]
\includegraphics[scale=0.6]{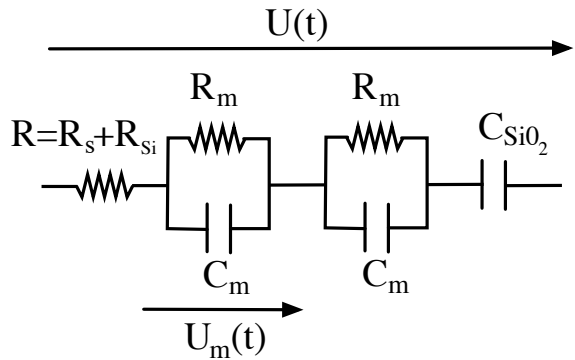}
\caption{A simple equivalent circuit for the system.}
\label{fig:6}
\end{figure}

\begin{figure}[h]
\includegraphics{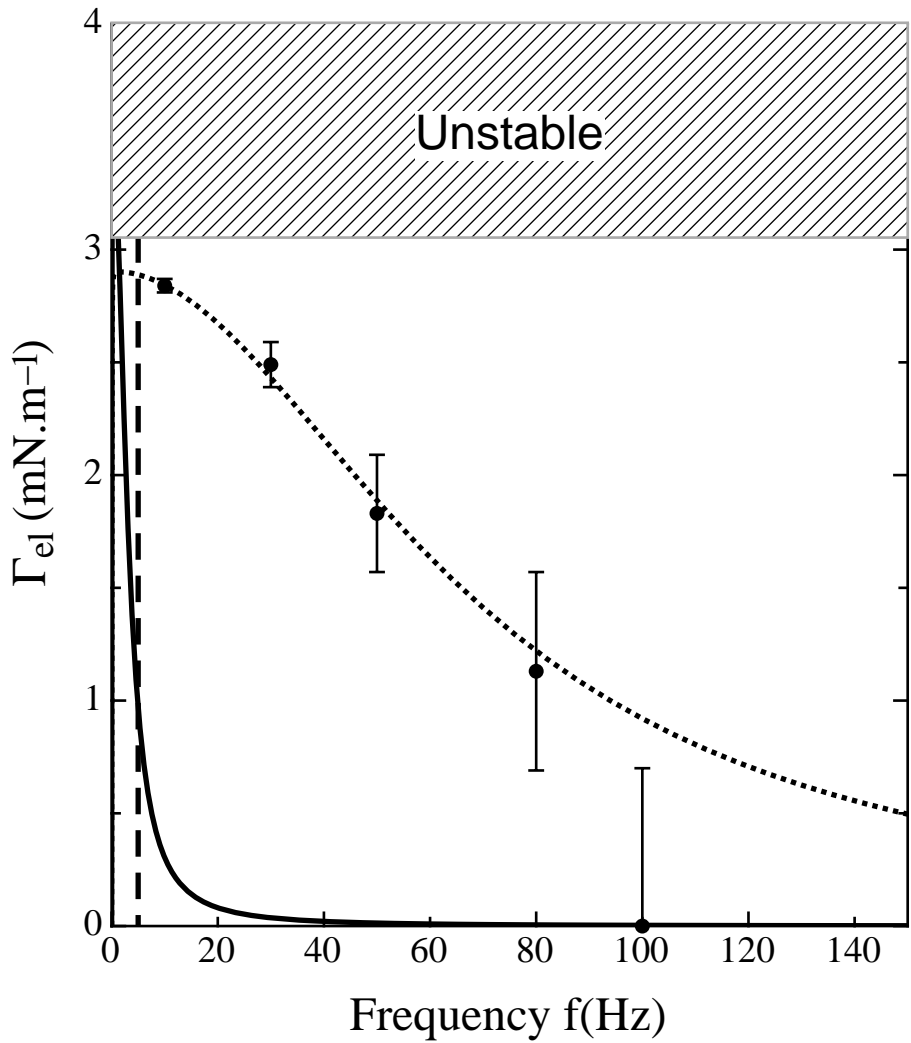}
\caption{$\Gamma_{el}$ as a function of frequency: ($\bullet$) values derived from experimental data; dashed line = best fit using equation \protect\ref{gammael=f(hz)}; full line = equation \protect\ref{gammael=f(hz)} with R$_m=60$ k$\Omega$ and C$_m = 2 \mu$F. We have represented by a dashed line the frequency limit (f $\sim 5$ Hz) were silicon oxidation is observed.}
\label{fig:7}
\end{figure}

Field frequency seems to be an important parameter in membrane destabilisation. We were not able to detect any modification (reversible or irreversible) of the bilayer above 100 Hz whereas the effect becomes important below 10 Hz (Fig. \ref{fig:4} and \ref{fig:7}). This is in good agreement with previous observations by Constantin et al. \cite{salditt2005}. It is also precisely the frequency range used in electroformation techniques. In order to understand more quantitatively this effect, let us try to estimate the local electric field acting on the bilayer. We can do so by using a simple equivalent circuit to represent the experimental set-up (figure \ref{fig:6}).

 The bilayer impedance ${\cal Z}_m$ is modelled by a resistance R$_m$ and a capacitance C$_{m}$ in parallel \cite{purrucker(2001)}. The substrate is described by a resistance R$_{Si}$ ($\ll 1 \Omega$) and a capacitance attributed to the silicon oxide C$_{Si0_2}$ ($\simeq$ 30 $\mu F$ for a 1 nm-thick oxide) in series. The electrolyte solution has a resistance $R_s$ ($\simeq 92$ k$\Omega$ for D$_2$O).  It is then easy to derive the negative electrostatic surface tension acting on the bilayer:
\begin{eqnarray}
\nonumber
\Gamma_{el} = \frac 12 \epsilon_m \left( \frac{R_m}{R} \right)^2 \frac{U_0^2}{\delta} \frac{\left( \omega \tau_i \right)^2}{\left( 1- \omega^2 \tau_m \tau_i \right)^2 + \omega^2 \left( \tau_m + \tau_i \left( 1 + n \beta \right) \right)^2} 
\label{gammael=f(hz)}
\end{eqnarray} 
where n is the number of bilayer(s), R the total resistance $R=R_s+R_{Si}$, $\beta = \frac{R_m}{R}$ and we have introduced the two characteristic times $\tau_m = R_m C_m$ and $\tau_i = R C_{SiO_2}$.

Using R$_s$ = 92 k$\Omega$ and C$_{SiO_2} = 30~\mu$F  as fixed values, the best fit to the data gives R$_m=48 \pm 1$ k$\Omega$ ($\sim 0.78 \pm 0.2$ M$\Omega$.cm$^{2}$) and C$_m = 0.1 \pm 0.01~\mu$F ($\sim 0.01 \pm 0.005~\mu$F.cm$^{-2}$).  The corresponding result is plotted figure \ref{fig:7}. The value of R$_m$ is in very good agreement with typical values of bilayer membranes resistance, around 0.9-1 M$\Omega$.cm$^{2}$ for a high coverage of the substrate  (see for example Ref. \cite{purrucker(2001),atanasov(2005)}). However the value of C$_m$ is surprisingly lower than values usually found for supported bilayers ($\sim 0.5~\mu$F.cm$^{-2}$ \cite{purrucker(2001),atanasov(2005)}). We have no explanation for this disagreement. Many effects could account for it, including the fact that the first membrane might be damaged by the field. Maybe our model is then too simple to give a good description of the system capacitive behaviour. We have drawn on figure \ref{fig:7} the curve obtained with a more consistent value of the membrane capacitance. In any case, we can notice that the frequency behaviour we observe is similar to previous observations by Constantin and coworkers \cite{salditt2005}.

\section{Conclusions and perspectives}

In this work we have reported the effect of an alternative electric field on lipid double bilayers adsorbed on doped silicon substrates, as studied by neutron reflectivity. We clearly show that the electric field can lead to a complete unbinding of the floating bilayer in a high conductivity solution at low frequency. In a low conductivity solution, small reversible modifications of the reflectivity profiles are observed and described as changes of the rms roughness of the floating bilayer. These data could be interpreted with a fluctuation spectrum including a negative electrostatic surface tension term as suggested by ref. \cite{sens2002}. 
Electric field frequency seems to be an important parameter as no effect is observed above 100 Hz. This result is also consistent with previous observations \cite{salditt2005}. We have tried to understand this effect with a simple electrokinetic model of the cell.\\

To the author's knowledge, this is the first time that the effect of an alternative electric field on the fluctuations of a membrane close to a wall is studied.  We have clearly shown that it is possible to completly destabilise a single bilayer by using an electric field in a well defined geometry. This work should be useful in the future for a better understanding of phenomena such as vesicle electroformation. \\
However, so far our results provide no information on the lateral length scale ($q^\star$ in ref. \cite{sens2002}) of the instability, which is one of the main predictions of the theoretical model \cite{sens2002}. We only measure the integrated root-mean-square amplitude $\sigma$ of fluctuations. In a near futur, we plan to perform off-specular reflectivity experiments, as described in ref. \cite{daillant2005}. They could provide important information concerning the lateral length scale(s) of the changes involved, and will enable direct measurement of the fluctuation spectrum and the surface tension.

Acknowledgments: The authors wish to thank Michel Bonnaud for technical assistance and the ILL for providing neutron beam-time. 
 
\newpage
 
\begin{table}[htpb]
\caption{Structural parameter of a DSPC double bilayer in D$_2$O before any electric field is applied: (sld) = scattering length densities in 10$^{-6}$ \AA$^{-2}$; thicknesses and roughnesses in \AA. All values are obtained from the best fit (figure \protect\ref{fig:2}).}
\begin{tabular}{|c|c|c|c c c|c|c c c|}
\hline
 & Silicon oxide & Solvent & Head & Chains & Head & Solvent & Head & Chains & Head\\
\hline 
 thickness & 11 $\pm$ 1 & 3 $\pm$ 0.5  & 6 $\pm$ 1 & 36 $\pm$ 1 & 6 $\pm$ 1& 19 $\pm$ 0.5 & 6 $\pm$ 1 & 36 $\pm$ 1 & 6 $\pm$ 1 \\
 sld & 3.4 $\pm$ 0.1 & 0 & 2.05 $\pm$ 0.1 & -0.6 $\pm$ 0.05 & 2.05 $\pm$ 0.1 & 0 & 2.05 $\pm$ 0.1 & -0.6 $\pm$ 0.05 & 2.05 $\pm$ 0.1\\
 \% solvent & 1 $\pm$ 1 & 100 & 19 $\pm$ 3 & 2 $\pm$ 1 & 19 $\pm$ 3 & 100  & 19 $\pm$ 3 & 2 $\pm$ 1 & 19 $\pm$ 3 \\ 
 roughness & 3 $\pm$ 0.5 & 2 $\pm$ 1 & 3 $\pm$ 1 & 4 $\pm$ 1 & 3 $\pm$ 1 & 5 $\pm$ 1 & 3 $\pm$ 1 & 6 $\pm$ 1 & 3 $\pm$ 1\\
\hline
\end{tabular}
\label{table1}
\end{table}%

%
%% For  wide figures use
%\begin{figure}
%% Use the relevant command for your figure-insertion program
%% to insert the figure file.
%% For example, with the option graphics use
%\resizebox{0.75\columnwidth}{!}{%
%  \includegraphics{leer.eps}
%}
%% If not, use
%%\vspace{5cm}       % Give the correct figure height in cm
%\end{figure}
%\begin{figure*}
%% Use the relevant command for your figure-insertion program
%% to insert the figure file. See example above.
%% If not, use
%\vspace*{5cm}       % Give the correct figure height in cm
%\end{figure*}
%\begin{table}
%\begin{tabular}{lll}
%\hline\noalign{\smallskip}
%\end{tabular}
%% Or use
%\vspace*{5cm}  % with the correct table height
%\end{table}
%
% BibTeX users please use
% \bibliographystyle{}
% \bibliography{}
%

\newpage

\bibliographystyle{epj}
\bibliography{/Users/thierrycharitat/Recherche/Membranes/bibliomembranes.bib}
\newcommand{\SortNoop}[1]{}

%\begin{thebibliography}{}
%Author, \textit{Book title} (Publisher, place year) page numbers
% etc
%\end{thebibliography}

%\newpage

\end{document}